\documentstyle[prl,aps,graphicx,amssymb,epsfig]{revtex}
\newcommand{\T}{${\mathcal T}_{3}$ }
\begin{document}

\twocolumn[
\hsize\textwidth\columnwidth\hsize\csname@twocolumnfalse\endcsname

\title{First observation of Aharonov-Bohm cages in 2-D normal metal networks}
\author{C\' ecile Naud, Giancarlo Faini, Dominique Mailly~$^*$}
\address{ L2M-CNRS, 196 Avenue H. Rav\'era, BP107, 92225 Bagneux Cedex, France \\
$*$e-mail:dominique.mailly@L2M.CNRS.fr}

\maketitle
\begin{abstract}
We report on magnetoresistance transport measurements performed on a bipartite tiling of rhombus in the GaAs/GaAlAs system. We observe for the first time large amplitude $h/e$ oscillations in this network as compared to the one measured in square lattices of similar size. These oscillations are the signature of a recently predicted localization phenomenon induced by Aharonov-Bohm interferences in this peculiar network.
\end{abstract}

\pacs{PACS numbers:7215.Rn,7215.Gd,7215.-v,7220.My,7320.Dx,7323.-b}

]

A magnetic field applied to a two-dimensional regular network induces a competition between the spatial period of the lattice and the magnetic period. During the last decades, this phenomenon of frustration induced by a magnetic field has been the object of intensive investigations. From a theoretical point of view, this subtle interplay leads to a complex energy spectrum which can be analysed using a simple tight-binding model for non-interacting electrons. For the square lattice case, the spectrum is described by the famous Hofstadter butterfly~\cite{Hofstadter}. A great number of geometries have been studied like, for instance, honeycomb~\cite{Rammal} and triangular lattices~\cite{Claro} but also fractal structures~\cite{jsaispas}. 
As phase coherence of the wave function is an essential ingredient, the first experiments have been performed in a network made of superconducting wires. In this case, there is a direct mapping between the solution of the linear Ginzburg-Landau equation and the tight-binding problem, leading to a direct link between the critical temperature and the energy spectrum \cite{Alexander,Pannetier}.
 
In the case of normal networks, the disorder plays an important role. For instance the Aharonov-Bohm (AB) effect, which modulates the interference pattern between two paths encircling a flux, is averaged to zero when dealing with an array of loops. This is due to the stochastic nature of the phase of the wave function in each individual cell, which depends on the microscopic configuration of the disorder. Thus, the $\Phi_{0}=h/e$ period signal decreases as $1/\sqrt{N}$ when $N$ loops are connected in series~\cite{umbach}. The only phase coherent effect which persists is the Altshuler-Aharonov-Spivak (AAS) oscillation~\cite{AAS} whose period is $h/2e$. In this case, interference concerns pairs of time reversal trajectories with a well defined phase which is modulated by the magnetic field. Then, all loops add coherently. In real samples, due to the non-zero width of the wires, there is a mixing of the different trajectories with various frequencies and phases, and only few oscillations are visible.

\begin{figure}[p!]
\centering
\includegraphics*[angle=-90,width=70mm]{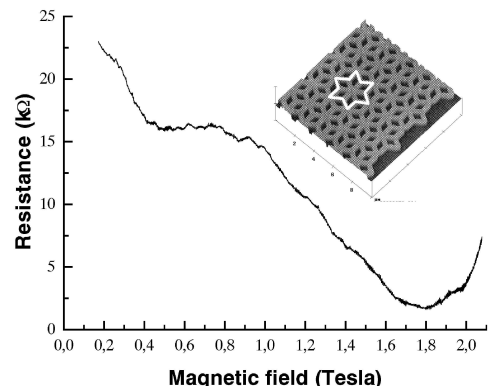}
\caption{Magnetoresistance at $T=30mK$ of a \T lattice between 0 and 2T. The inset shows an AFM view of the sample. The width of each wire is about 0.4$\mu m$ and the area of a unit rhombus is equal to $0.8\mu m^{2}$, leading to a quantum flux $\Phi_{0}$ for $B=50$Gauss. An AB cage is underlined in white}
\label{fig:RH}
\end{figure}

Recently, a new localization phenomenon induced by a magnetic field in a 2D lattice has been reported~\cite{vidal}. The authors consider a bipartite hexagonal structure containing three sites per unit cell, one sixfold coordinated and two threefold coordinated. In this so-called \T lattice, the propagation of the electron wave function is bounded in a small number of cells, called an AB cage, for a particular magnetic flux ($\Phi=\Phi_{0}/2$, see the AFM picture of our sample in the inset of fig.~\ref{fig:RH}). The tight-binding model shows that, for this magnetic flux, the energy spectrum collapses into three infinitely degenerate levels, predicting a divergence of the resistance. The overall spectrum is periodic with a period $h/e$.

This theoretical result is obtained for a pure network made of sites coupled by hopping. One may ask wether this effect will persist in a network made of 1D metal wires. This point, as well as the effect of the disorder, have been recently investigated theoretically~\cite{montambaux}. The cage effect is expected to be more robust against disorder averaging than the standard AB oscillations. This strong localization effect should hence be observable in transport measurements by periodic modulation of the resistance, with a period $h/e$.

The first experiments on the \T lattice have been performed on a superconducting material~\cite{abilio}. The bottom of the computed energy spectrum has been reproduced through the variation of the critical temperature versus the magnetic flux. As far as the localization is concerned, it has been evidenced by a strong depression of the critical current and a broadening of the superconducting transition at $\Phi=\Phi_{0}/2$. Up to now, no measurement has been reported on normal metal transport properties. 

In this letter, we present the first evidence of AB cages in a normal network tailored in a high mobility two dimensional electron gas (2DEG).
 Low temperature magnetoresistance measurements of \T lattices show clear $h/e$ oscillations in arrays of 2500 cells. From the temperature dependence of these oscillations we are able to extract a large characteristic length compared to that measured on a single cell. Experiments performed on square lattices of similar size do not show such a behavior, further supporting the existence of the cage effect. More strikingly, at high magnetic field, $h/2e$ oscillations appear whose amplitudes can be much higher than the fundamental period. The temperature dependence is similar to that of the $h/e$ signal. These observations dismiss a simple interpretation in terms of harmonics generation. The origin of this phenomenon is still unclear and needs more investigations.

We have used MBE grown AlGaAs-GaAs heterojunctions as starting material to fabricate the samples. When cooled down to liquid Helium temperature the as-grown 2DEG has an electron density $n=3~10^{11}cm^{-2}$ and a mobility $\mu=10^{6}cm^{2}V^{-1}s^{-1}$. These values yield a Fermi wavelength $\lambda_{f}=65nm$ and an elastic mean free path $l_{e}=6.4\mu m$. With these parameters, the phase coherence length is usually of the order of $L_{\Phi}=20\mu m$ below T=0.1K~\cite{cp}. A JEOL 5DIIU electron beam writer was used to define the sample patterns. By lift-off, we first deposited an aluminum mask which was subsequently transfered into the 2DEG by Argon ion etching. The inset of figure~\ref{fig:RH} shows a detailed view of the \T tiling. Between the voltage probes, the array spans a $80\mu m \times 25\mu m$ surface. Since the area of a unit rhombus cell is $0.8\mu m^{2}$ we probe about 2500 cells, one flux quantum $\Phi_{0}$ per unit cell corresponding to a magnetic field of $50G$.

The nominal width of the wire defining the network is $0.4\mu m$, but the effective electrical width is considerably smaller due to the lateral depletion resulting from the etching process. Indeed, by varying the energy of the ions and/or the etching time, we can prepare samples presenting a slight side depletion of about $0.1\mu m$ to a completely depleted wire. This is an important point since in the Quantum Hall Effect (QHE) regime, the current is carried by edge states: opposite sample edges carrying the flow in opposite directions~\cite{Buttiker}. In a ring geometry only the inner states encircle a flux and the AB signal vanishes for wide enough samples~\cite{timp}. As in high mobility materials the appearance of the QHE regime starts at rather small magnetic fields (less than 1000Gauss), the real width of the samples must be small enough if one wants to observe oscillations for a comfortable range of magnetic field. In order to investigate the effect of lateral depletion, we have then fabricated sets of wires of different widths. When the conductance at $T=4K$ is no longer proportional to the mask width, we conclude that only few conducting channels are present. Thus we adjust the etching parameters to fulfill this criterion for a mask wire width of $0.4\mu m$. Indeed, with somewhat shallow etching, well defined Shubnikov-de-Haas (SdH) oscillations are observed at low temperature on our arrays indicating the presence of edge states. Increasing the etching time and/or energy washes out these oscillations.We have investigated about ten samples prepared with the above described deep etching process. The square resistance of the arrays was always below $10k\Omega$ leading to a resistance per connecting linesmaller than $12k\Omega$, charging effects are thus not relevant. This is confirmed by linear I-V characteristics. All samples made this way have shown $h/e$ signal.

The samples are placed on the cold finger of a dilution refrigerator and cooled down to 30mK. A perpendicular magnetic field of up to 7T can be applied. Measurements are recorded using an ac resistance bridge working at 33Hz or lockin amplifiers with a current injection of 10nA.

\begin{figure}
\centering
\includegraphics*[width=70mm]{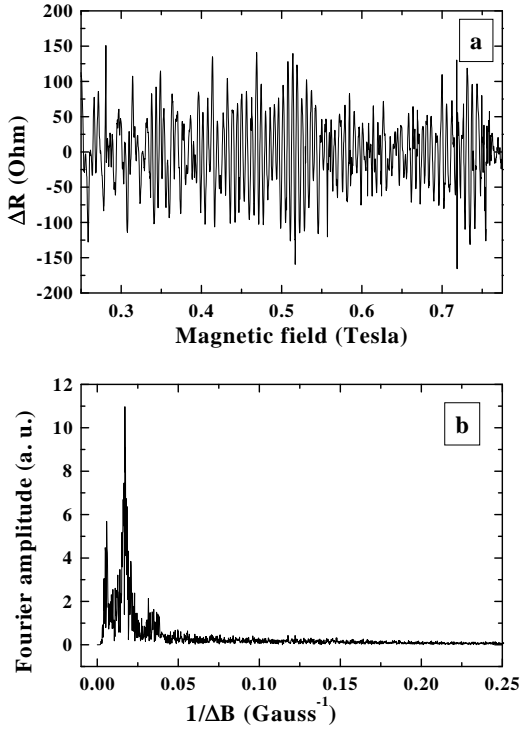}
\caption{\textbf{a)} Amplitude of the small period oscillations for the \T lattice versus the magnetic field, extracted from the experimental data using the procedure described in the text. \textbf{b)} Fourier transform of the above oscillations. A clear peak appears at $1/55 G^{-1}$ corresponding to a flux quantum $h/e$ in a unit cell.}
\label{fig:TFS}
\end{figure}

A typical magnetoresistance plot for a \T lattice is shown in figure~\ref{fig:RH}. Rapid oscillations are superimposed on large scale fluctuations of the overall signal. We substract a polynomial fit of the data to the experimental points in order to extract the small period oscillations. This resulting signal, shown in  fig.~\ref{fig:TFS}a, is then Fourier transformed (fig.~\ref{fig:TFS}b). A clear peak appears on the spectrum at a frequency corresponding to 55G, the width of which is $5$G. This width arises due to the flux difference between the most inner and the most outer trajectories in a rhombus. From this measured value we can estimate the effective wire width to be $20$nm. The typical amplitude of the oscillations is $\Delta R = 100 \Omega$ leading to a value of $0.02 e^{2} / h$ in terms of conductance. This has to be compared to the one obtained from a single rhombus. We measured a $0.05 e^{2} / h$ typical amplitude for a single unit tile fabricated on the same dice. The expected amplitude reduction due to the above discussed averaging process ($1/\sqrt{2500}$) would then yield  an amplitude of $0.001 e^{2} / h$. This is far less than the one we measure, giving a first indication of the presence of cages for the \T lattice.

\begin{figure}
\centering
\includegraphics*[width=70mm]{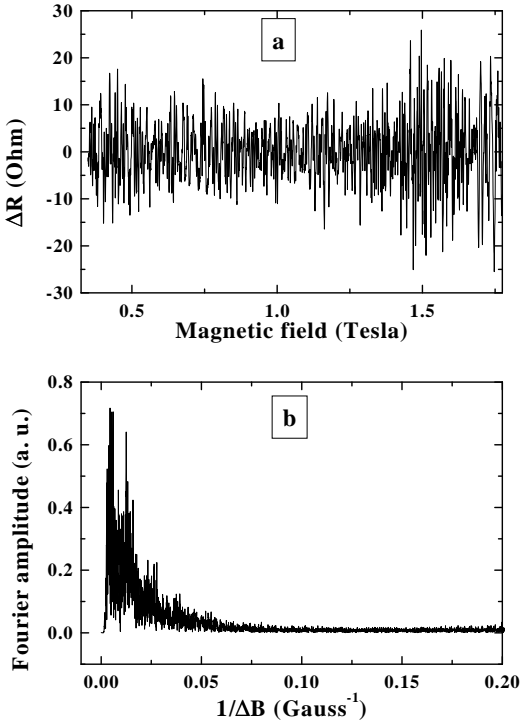}
\caption{\textbf{a)} Same than in fig~\ref{fig:TFS}a) but for the square lattice. Note that the amplitude in this case is about ten times smaller than for the \T lattice. \textbf{b)} Fourier transform of the above signal.}  
\label{fig:TFC}
\end{figure}

To validate the peculiarity of this topology, for each etching run we have also patterned on the same dice $80\mu m \times 25\mu m$ square lattices with similar wire width and unit cell area. The typical square resistance of these arrays is in the same range than that of the \T network. Following the same data analysis, we obtain typically in this case the features shown in fig.~\ref{fig:TFC}. One can hardly distinguish a peak in the Fourier transform signal. The $h / e$ signal, if any, is embedded in the $1/f$ universal fluctuations contribution, its amplitude being at least one order of magnitude smaller than the \T one for all the investigated samples.

\begin{figure}
\centering
\includegraphics*[width=70mm]{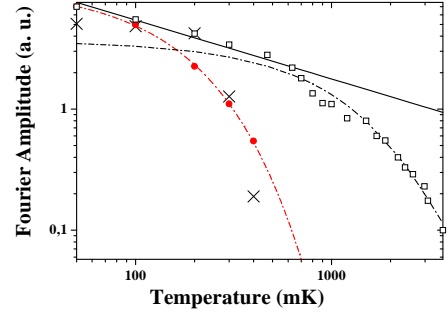}
\caption{Temperature dependence  of the peak amplitude of the Fourier transform normalized to the value at $T=50mK$. $(\bullet)$ and $(\times)$ refer to the $h/e$ and $h/2e$ signals  respectively for the \T lattice. ($\square$) refers to the $h/e$ signal for the single rhombus. Solid line is the $T^{-1/2}$ fit and dotted line the exponential fits.}
\label{fig:TFT}
\end{figure}

In fig.~\ref{fig:TFT} we have plotted on a log-log scale the temperature dependence of the $h / e$ peak of the  Fourier transform for both the \T lattice ($\bullet$) and the single rhombus ($\square$). For a single loop the amplitude of the AB signal is expected to follow a $T^{-1/2}$ law, because of temperature averaging, as long as the size of the loop is smaller than $L_{\Phi}$. When the temperature is such that $L>L_{\Phi}$ the AB signal falls off exponentially~\cite{milliken}. Then, we expect this critical temperature to be lower for the \T network compared to that of a single tile, since for the former the characteristic size is the perimeter of a cage, namely $12\mu m$, to be compare to the 4$\mu m$ perimeter of a rhombus (see inset of figure~\ref{fig:RH}). The experimental results depicted in fig.~\ref{fig:TFT} show that this cut-off temperature is around 1K for the rhombus whereas it is below 100mK for the \T. This is another important evidence of the role of the AB cages in the transport properties of this peculiar network.

\begin{figure}
\centering
\includegraphics*[width=70mm]{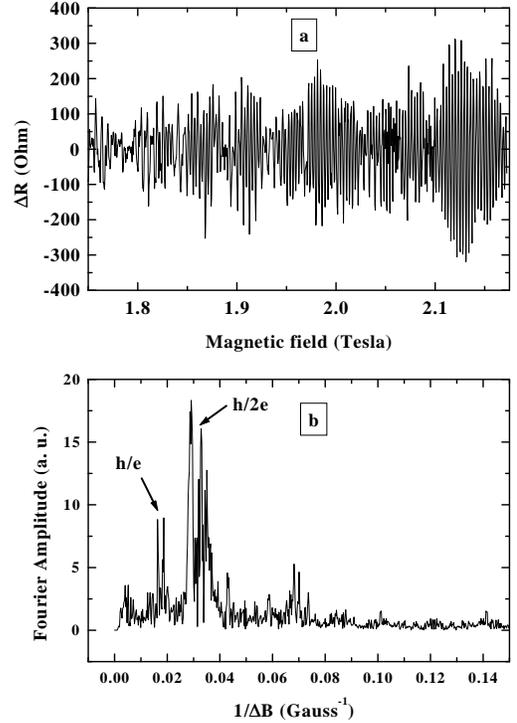}
\caption{\textbf{a)} Amplitude of the small period oscillations for the \T lattice extracted from the experimental data in the strong magnetic field range. \textbf{b)} Fourier transform of the above signal.  The $h/2e$ peak is much larger than that of the fundamental. }
\label{fig:TFT2}
\end{figure}

Finally, in the high magnetic field regime we observe a striking  frequency doubling in the spectra. If this signal were a simple harmonic, one would expect the $h/2e$ peak to be much smaller than the $h/e$ peak since it corresponds to a trajectory twice as long~\cite{washburn}. But the amplitude of this peak is of the same order or even larger than that of the fundamental. This is shown in figure~\ref{fig:TFT2} where the relative amplitude variation of the magneto-resistance is plotted as well as its Fourier transform from 1.5 to 2.17 T. Moreover, if this signal were a simple harmonic, the characteristic temperature should be smaller. This is in contradiction with the measured temperature dependence of the $h/2e$ peak amplitude ($\times$ in fig.~\ref{fig:TFT}) which shows a similar decay as that observed for the fundamental period. It is worth noting that we never observed such a behavior in the square lattices. Up to now, we have no hint on the nature of this frequency doubling. Let us also stress that this signal appears for strong magnetic fields where the magneto-resistance presents important variations. Of course, one may think of AAS oscillations  when dealing with an $h/2e$ signal. As mentioned in the introduction, due to the aspect ratio of the sample, the AAS signal should vanish for high magnetic fields. Thus, an explanation in term of AAS would imply that a high magnetic field induces the squeezing of the wires. But in this case, one expects to observe a similar effect for the square lattice, unless the geometry of the \T network again favors the appearance of the $h/2e$ oscillations. The study of different types of networks and the addition of a gate to control the depletion is under progress to clarify this phenomenon.

In conclusion, we have measured for the first time large Aharonov-Bohm type oscillations in an array of rhombus. The amplitude detected and its temperature dependence are in favor of a recently suggested cage effect. The comparison with a square lattice unambigously indicates that this behavior is related to the topology of this peculiar network. We also observe unexpected high amplitude $h/2e$ oscillations which are still not understood. Nevertheless, this \T type of network, opens the way to new experimental systems where the great number of unit cells probed wash rules out the problem of reproducibility usually encountered when dealing with a unique small object. With the AlGaAs-GaAs system we used, we are very close to the model of a network of wires with a small number of channels used in numerical calculation. We can then easily investigate samples with controlled topological defects or disorder and compare their behavior with theory.

We are endepted to B. Etienne and A. Cavanna for supplying high quality 2DEG. We kindly acknoledge fruitful discussions with H.~Bouchiat, B.~Dou\c cot, G.~Montambaux, R.~Mosseri, J.~Vidal and A. Ramdane for critical reading of the manuscript. We benefit from the technical skills of L.~Couraud, X.~Lafosse, A.~Cavanna and C.~David for the AFM picture. This work is supported by the R\'egion Ile-de-France through a SESAME contract.

\begin{sloppypar}

\end{sloppypar}


\end{document}